\RequirePackage{fix-cm}
\documentclass[twocolumn,epjc3]{svjour3}  
\smartqed  
\RequirePackage{graphicx}
\usepackage{fancybox}
\usepackage{ascmac}
\usepackage{xcolor}
\usepackage{subfigure}
\usepackage{here}
\usepackage{mathtools}
\usepackage{amssymb}
\PassOptionsToPackage{hyphens}{url}
\usepackage{hyperref}
\usepackage{cuted}
\usepackage{lineno}
\journalname{Eur. Phys. J. C}
\begin{document}

\title{Constraining the strangeness enhancement scenario of the UHECR muon puzzle with LHC experiments}

\author{Ken Ohashi\thanksref{e1,addr1}
        \and
        Anatoli Fedynitch\thanksref{addr2} 
        \and 
        Hiroaki Menjo\thanksref{addr3,addr4}
}
\thankstext{e1}{e-mail: ken.ohashi@unibe.ch}

\institute{Laboratory for High Energy Physics, Albert Einstein Center for Fundamental Physics, University of Bern, Bern, Switzerland\label{addr1}
\and High-Energy Theory Group, Institute of Physics, Academia Sinica, Taipei, Taiwan \label{addr2}
\and
Institute for Space-Earth Environmental Research, Nagoya University, Nagoya, Japan \label{addr3}
\and 
Kobayashi-Maskawa Institute for the Origin of Particles and the Universe, Nagoya University, Nagoya, Japan
\label{addr4}
}
\date{Received: date / Accepted: date}

\maketitle

\abstract{
The excess of muons observed in ultra-high-energy cosmic-ray air showers relative to simulation predictions, known as the muon puzzle, provides indirect evidence of our incomplete understanding of high-energy hadronic interactions. An unambiguous resolution requires that each proposed solution be directly tested through cosmic-ray and collider experiments probing hadronic interactions. In this work, we develop a framework to assess the strangeness enhancement scenario, wherein an increased yield of kaons relative to pions boosts muon production, which can connect a model prediction and cosmic-ray and collider measurements. Using the \textsc{MCEq} air-shower simulation package, we first identify the key phase-space regions of hadronic interactions that drive muon yields in this scenario. 
Subsequent analysis demonstrates that a strangeness enhancement starting at $10^6-10^7~\mathrm{GeV}$ can consistently explain the latest cosmic-ray experiments and requires substantial enhancement at the Large Hadron Collider (LHC) energy.  
Furthermore, evaluating the required precision for LHC measurements, assuming Pierre Auger Observatory muon measurements and forthcoming kaon-to-pion ratio data from LHC Run~3, reveals that these experiments can robustly constrain the majority of the scenario's parameters. In particular, achieving 10.8\% precision on the kaon-to-pion ratio at LHCb and 8.4\% at FASER is sufficient to test the strangeness enhancement scenario over its viable parameter space. These upcoming experimental results will provide the first direct constraints on strangeness enhancement as a potential resolution of the muon puzzle.
}


\section{Introduction}

Ultra-high-energy cosmic rays (UHECRs) are the most energetic particles observed, primarily consisting of protons or atomic nuclei from unidentified astrophysical sources. Upon reaching Earth, they interact with atmospheric nuclei and generate extensive air showers composed of secondary particles. Large-scale observatories such as the Pierre Auger Observatory (PAO) and the Telescope Array experiment measure secondary particles at ground level. A persistent excess of muons compared to simulations has been reported by several experiments, including PAO~\cite{MuonAuger2015,PierreAuger:2016nfk,WHISP2023,PierreAuger:2024neu}, particularly at primary energies above $10^8$\,GeV, a discrepancy commonly referred to as the “muon puzzle”~\cite{Albrecht:2021cxw}.

A cosmic-ray-induced air shower consists of a mixture of electromagnetic and hadronic components. The electromagnetic component originates mainly from the decay of secondary $\pi^0$ mesons, while the hadronic cascade is driven by protons, $\pi^\pm$, and $K^\pm$. Muons primarily arise from the decay of low-energy $\pi^\pm$ mesons near the end of the cascade. Simulations fail to simultaneously reproduce both components accurately due in part to our limited understanding of hadronic interactions. First-principles QCD calculations are not applicable at low momentum transfers, necessitating the use of phenomenological models. Studies have shown that the energy-sharing ratio between neutral pions and other hadrons is a key parameter for adjusting muon yields with limited impact on other observables~\cite{Pierog:2006qv,Drescher:2007hc,Baur:2019cpv,Ostapchenko:2024myl,Riehn:2024prp,Ulrich:2010rg,Albrecht:2025kbb}.

Several hypotheses have been proposed to explain the muon excess. One is \emph{strangeness enhancement}~\cite{Baur:2019cpv,Anchordoqui:2016oxy,Manshanden:2022hgf,Anchordoqui:2022fpn,Scaria:2023coa}, where kaons are produced more abundantly at the expense of pions. This reduces the energy feeding the $\pi^0$-initiated electromagnetic component and increases muon production. The idea is inspired by ALICE measurements of strange hadron production in high-multiplicity events~\cite{ALICE:2016fzo}. 
Although these results pertain to low-energy secondaries at mid-rapidity, a similar mechanism at forward rapidities and high energies could help explain the muon puzzle. 
Another hypothesis is \emph{$\rho^0$ enhancement}, in which $\rho^0$ mesons replace $\pi^0$ mesons in high-energy interactions. These ideas have been tested in simulations by modifying hadronic final states, for example in AIRES~\cite{Anchordoqui:2022fpn} or in SIBYLL2.3 via the SIBYLL$^\bigstar$ scenarios\cite{Riehn:2024prp}, both of which reproduced the observed muon counts under certain assumptions.

Previous studies of strangeness enhancement have primarily focused on demonstrating that such mechanisms can, in principle, explain the muon excess when model parameters are chosen a priori~\cite{Baur:2019cpv,Anchordoqui:2022fpn,Riehn:2024prp}. However, these studies did not address a critical question: how can cosmic-ray and accelerator experiments quantitatively test these scenarios? 
Specifically, it remains unclear where in phase space measurements should be made, what precision is required, and which observables are most effective for constraining or rejecting the enhancement hypothesis? This work addresses these questions by providing a quantitative framework that bridges the gap between cosmic-ray observations and accelerator-based tests of strangeness enhancement.

Understanding whether the muon excess stems from hadronic mismodeling requires supporting evidence from accelerator experiments. The Large Hadron Collider (LHC), the highest-energy proton and ion collider, offers valuable constraints on particle production. However, three of its major experiments, ATLAS, CMS, and ALICE, are optimized for mid-rapidity, and the collider’s center-of-mass energy remains below that of UHECR interactions. Several experiments, LHCb, LHCf, FASER, and SND@LHC, measure hadron production in the forward-rapidity region.
Nonetheless, air-shower studies suggest that small discrepancies in hadronic models can be amplified through cascading secondary interactions~\cite{PierreAuger:2021qsd}. A moderate increase in the charged-hadron-to-$\pi^0$ ratio, particularly through enhanced forward kaon production, could thus account for the muon excess. Notably, $K^0$ mesons do not decay purely electromagnetically, so an increased $K^0/\pi^0$ ratio would reduce energy transfer to the electromagnetic channel.

In this work, we develop a framework to assess the parameter space of a muon-puzzle scenario and connect model predictions with present and future experimental constraints. First, the framework quantifies the relevance of hadronic interactions by implementing strangeness enhancement in simulations. Subsequently, our analysis identifies regions of parameter space within the scenario that are consistent with the latest cosmic-ray observations. Finally, the study evaluates the precision required for LHC measurements to probe or constrain the scenario parameters. Sect.~\ref{sec:calculation_for_contour} introduces the calculation method for interaction importance, and Sect.~\ref{sec:results_contour} presents the results. Sect.~\ref{sec:method_for_comparison} and Sect.~\ref{sec:results_constraints} detail and apply the method for assessing the strangeness enhancement scenario. Conclusions follow in Sect.~\ref{sec:conclusion}.

\section{Method for Estimating Interaction Importance\label{sec:calculation_for_contour}}

\subsection{Simulation of air showers under varying particle yields}
We use the \textsc{MCEq} simulation package~\cite{Fedynitch:2015zma,Fedynitch:2018cbl} together with the hadronic interaction model \textsc{SIBYLL}~2.3d~\cite{Riehn:2019jet}. The \textsc{MCEq} package uses inclusive particle yield tables arranged in matrices to solve the coupled transport equations for particle cascades in the Earth's atmosphere. Although the code is originally developed to calculate atmospheric lepton fluxes, it is straightforward to apply to compute average particle densities as a function of the depth in air showers initiated by a single primary particle.

In this work, we concentrate on modifying the secondary particle production yields in MCEq's $\mathbf{C}$ matrix. The matrix elements are binned inclusive secondary particle yields $c_{ij} = {\rm d}N(E^i_\text{proj}, E^j_\text{sec})/{\rm d}E_\text{sec}$. The kinetic energy grids are identical for projectiles and secondary particles and span the energy range of $\sim0.1$ GeV up to $10^{11}$~GeV, resulting in approximately $120\times120$ coefficients. Each ``interaction channel'', such as $\pi^+$ produced in proton-air interactions or $K^-$ produced by $\pi^+$-air interactions, is represented by an independent matrix, so \textsc{MCEq} consists of dozens of such matrices, which are arranged in a coupled equation system to be solved. By perturbing the phase space elements, e.g.~$c_{ij} \to c_{ij} (1 + \epsilon)$, regenerating the equation system, and recalculating the air shower observable, we establish a method to study the relevance of a particle production phase space element for all relevant energy scales.

\subsection{Implementation of strangeness enhancement}

We implement the strangeness enhancement scenario by substituting a fraction $\zeta$ of neutral and charged pions with kaons within a given phase space element $c \equiv c_{ij}$ by
\begin{align}
    c_{\pi^{\pm0}}^{\text{mod}} &= (1 - \zeta)\,c_{\pi^{\pm0}}, \label{eq:pi-mod} \\
    c_{K^{\pm0}}^{\text{mod}} &= c_{K^{\pm0}} + \zeta\, c_{\pi^{\pm0}}. \label{eq:K-mod}
\end{align}
A fraction $\zeta$ of the pion yield is reassigned to kaons (and correspondingly removed from pions) to preserve the total inclusive yield in that phase-space bin.
The matrix elements for $\pi^\pm$, $\pi^0$, $K^\pm$, $K^0_L$, and $K^0_S$ are modified simultaneously. For $\pi^0$, half of its yield is swapped with $K^0_L$ and $K^0_S$, and the sum of the $K^0_L$ and $K^0_S$ contributions is swapped into $\pi^0$.

We consider $p$, $\bar{p}$, $n$, $\bar{n}$, $\pi^+$, and $\pi^-$ projectiles, as these are the most abundant projectiles in air showers. For pion projectiles, a correction is included for leading particles. These are secondaries sharing at least one of the valence quarks with the projectile, carrying the highest energy fraction, and typically originating from the fragmentation of valence quarks. This generally is the same meson, i.e., a secondary $\pi^-$ from a $\pi^-$ projectile. To preserve these basic quark counting rules for charged pions, the modification in Eqs.~(\ref{eq:pi-mod}) and (\ref{eq:K-mod}) is only applied to the non-leading contribution, defined as $c_{\pi^\pm}^{\text{non-leading}} = c_{\pi^\pm} - c_{\pi^\mp}$.

\subsection{Calculating phase space gradients on air shower observables}

To compare with the muon yields measured by the PAO, we simulate air showers initiated by protons, helium, nitrogen, and iron nuclei at primary energies $10^8$~GeV, $10^{8.5}$~GeV, $10^{9.5}$~GeV, and $10^{10}$~GeV at a fixed zenith angle of $67^\circ$.

\noindent For a primary proton with energy $E_\text{prim}$, the \emph{gradient matrix} $\mathcal{D}$ is defined as
\begin{equation}
    \mathcal{D}(E_\text{prim}, T_{\text{proj}}, E_{\text{proj}}, x_{\text{lab}} ) = \frac{1}{N_\mu}\, \frac{\partial N_\mu}{\partial \zeta}\, .
    \label{eq:D-matrix}
\end{equation}
The arguments are projectile type ($T_{\text{proj}} = $ proton, neutron, $\pi^\pm$, and $K^\pm$), projectile energy ($E_{\text{proj}}$) and $x_{\text{lab}} = E_{\text{sec}} / E_{\text{proj}}$, where projectile refers to any hadron initiating an interaction, including the primary particle. The gradients are computed using a finite difference scheme, where each $c_{ij}$ of the yield matrices is individually perturbed with a small $\zeta = 0.005$ according to Eqs.~(\ref{eq:pi-mod}) and (\ref{eq:K-mod}) while excluding unphysical configurations $E_{\text{sec}} > E_{\text{proj}}$. The gradient $\partial N_\mu/\partial \zeta$ is computed as the relative change in the muon yield between the modified and unmodified matrices. We fix $\zeta = 0.005$ for these derivative evaluations and verify that, for $x_{\text{lab}} > 10^{-4}$, the derivative matrix $\mathcal{D}$ is insensitive to the exact value of $\zeta$, indicating a linear response in this regime. As explained in Sect.~\ref{sec:method_for_comparison}, the equivalent $\mathcal{D}$ for nuclei is obtained from a superposition approximation.

\section{Key Phase-Space Regions in the Muon Puzzle\label{sec:results_contour}}
\begin{figure*}
    \centering
    \includegraphics[width=0.95\textwidth]{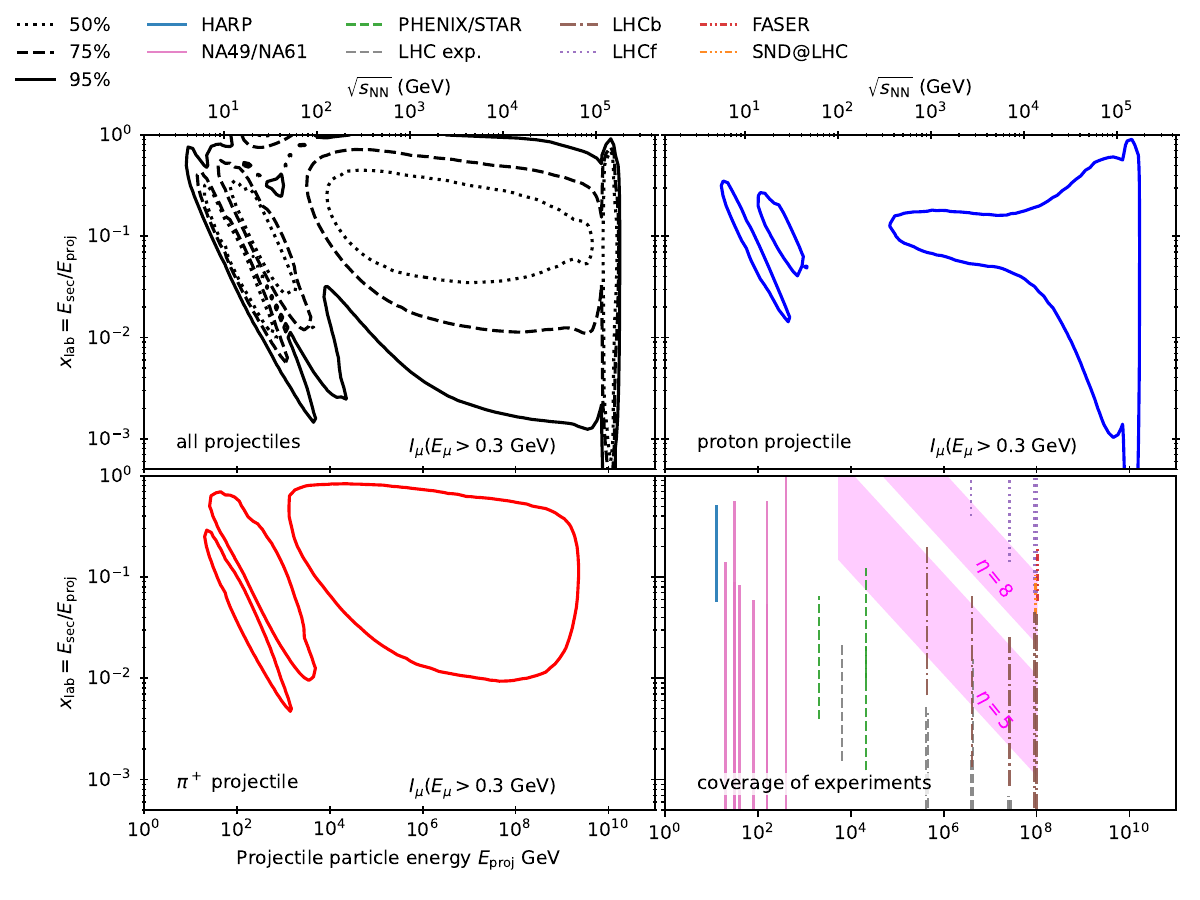}
    \caption{Contour plots showing regions enclosing 50\% (dotted), 75\% (dashed), and 95\% (solid) of the total effect for a $10^{10}$~GeV proton primary. 
    The left panel displays the combined response over all projectile types; the right and lower-left panels show the contributions from proton, neutron, and $\pi^+$ projectiles, respectively. The coverage of experiments is shown in the bottom right panel.
    The pseudorapidity acceptance of the collider experiment is shown in the hatched regions, assuming $p_T$ from 0.1 to 1.0~$\mathrm{GeV}$ for $\eta=5.0$ and $p_T$ from 0.1 to 0.5~$\mathrm{GeV}$ for $\eta=8.0$.}
    \label{fig:contours1}
\end{figure*}
\begin{figure*}
    \centering
    \includegraphics[width=0.95\textwidth]{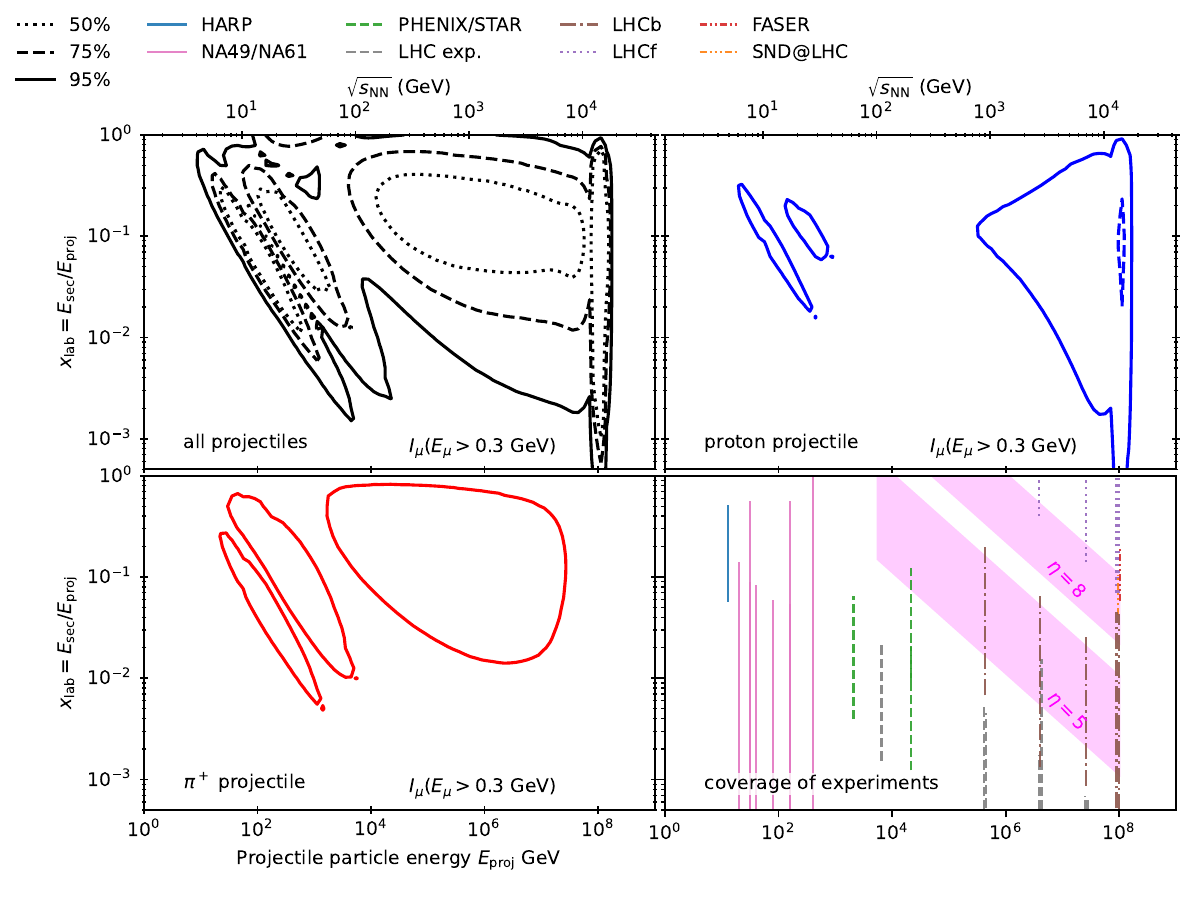}
    \caption{Contours enclosing 50\% (dotted), 75\% (dashed), and 95\% (solid) of the total effect for a $10^8$~GeV proton primary. The left panel displays the combined response over all projectile types; the right and lower-left panels show the contributions from proton, neutron, and $\pi^+$ projectiles, respectively. The coverage of experiments is shown in the bottom right panel.
    The pseudorapidity acceptance of the collider experiment is shown in the hatched regions, assuming $p_T$ from 0.1 to 1.0~$\mathrm{GeV}$ for $\eta=5.0$ and $p_T$ from 0.1 to 0.5~$\mathrm{GeV}$ for $\eta=8.0$. 
    }
    \label{fig:contours2}
\end{figure*}
We identified the phase-space regions in which interactions significantly affect the muon puzzle by visualizing the derivative matrix $\mathcal{D}$ defined above. We represented $\mathcal{D}$ with contours, where the fraction $C$ of significant interactions is defined by
\begin{equation}
    \frac{\int_{E_A} dE'\, |\mathcal{D}(E')|}{\int dE\, |\mathcal{D}|} = C\,,
\end{equation}
where $E_A$ denotes the region of phase space that encloses the fraction $C$. We consider $C = 0.50$, $0.75$, and $0.90$.

Figure~\ref{fig:contours1} shows the results for a $10^{10}$~GeV proton primary and three types of projectiles in an air shower. Three important phase-space regions emerge: (a) the interaction induced by the primary cosmic ray; (b) interactions of secondary projectiles with $10^4$~GeV $< E_{\text{proj}} < 10^{-2} E_{\text{prim}}$; and (c) interactions of secondary particles with $10^2$~GeV $\lesssim E_{\text{proj}} \lesssim 10^4$~GeV. In region~(a), particles with $x_{\text{lab}}$ down to $10^{-3}$ significantly affect the muon number during the first few interactions. In region~(b), only very forward secondaries with $10^{-2} < x_{\text{lab}} < 5\times 10^{-1}$ have a notable impact; all projectile types contribute here, with charged pion interactions dominating. In region~(c), low-energy baryon and meson subcascades with $E_{\text{sec}} \sim 10^2$~GeV produce most GeV-scale muons at ground, with the $x_{\text{lab}}$ range limited by the applied cutoff $E_\mu > 0.3$~GeV.

Figure~\ref{fig:contours2} presents the corresponding contours for a $10^8$~GeV proton primary. The qualitative behavior is consistent across primary energies: region~(c) continues to dominate muon production, while the relative importance of regions~(a) and~(b) shrinks with decreasing primary energy. Since a larger discrepancy in muon number is observed for $10^{10}$~GeV cosmic rays compared to $10^8$~GeV~\cite{WHISP2023}, this suggests that region~(c) alone cannot account for the muon puzzle. Instead, the results point to interactions in regions~(a) and~(b), those induced by hadrons with $E_{\text{proj}} > 10^4$~GeV and $x_{\text{lab}} > 10^{-3}$, as the critical contributors to resolving the discrepancy.

The phase space accessible to accelerator experiments is shown as vertical lines at discrete beam energies in the bottom right panel of Figs.~\ref{fig:contours1} and \ref{fig:contours2}. Solid and dashed lines indicate detector coverage in collider and fixed-target experiments, respectively. Converting $E_{\text{proj}}$ into the nucleon–nucleon center-of-mass energy $\sqrt{s_{NN}}$ allows us to map each experiment’s detector acceptance into $x_{\text{lab}}$ (see Appendix~A for details). Fixed-target experiments generally cover region~(c), while collider experiments partially cover region~(b). In particular, the LHCb, LHCf, FASER, and SND@LHC experiments~\cite{LHCbRun2,LHCfpi0_ICRC2021,LHCfpi0_2015,FASER_detector,SNDLHC:2022ihg} probe regions within the 75\% contour, because they cover the forward region ($\eta > 2.5$) and allow identification of pions and kaons. LHCf has reported results on photons, neutral pions, and neutrons, which are insufficient to directly constrain a kaon enhancement. 

While LHCf constrains the electromagnetic component by measuring the $\pi^0$ yield~\cite{LHCfpi0_ICRC2021,LHCfpi0_2015}, the experiment cannot directly measure charged pions and kaons that drive muon production, and thus cannot independently determine the electromagnetic-to-hadronic ratio, which is a key element for testing the strangeness enhancement scenario. Furthermore, LHCf's typical $x_{\rm lab}$ is approximately $0.2$, thereby missing the most critical region for muon production where the enhancement effects are most pronounced.

\section{Method for Comparison with Experimental Data\label{sec:method_for_comparison}}

Previous studies of strangeness enhancement scenarios have provided phenomenological benchmarks by evaluating discrete sets of model parameters. The present analysis utilizes the \textsc{MCEq} simulation package to compute air-shower observables with significantly reduced computational overhead. This efficiency permits a systematic evaluation of continuous parameter variations, extending previous single-point evaluations into a comprehensive mapping of the relevant parameter space.

\subsection{Implementation of the Strangeness Enhancement Scenario}
In this section, we outline a method to calculate the expected muon yield under a simplified strangeness-enhancement model and compare these predictions with current PAO results and projected data from LHC Run 3. The model assumes that strangeness enhancement in secondary interactions depends on the projectile energy $E_{\text{proj}}$ and the lab-frame energy fraction $x_{\text{lab}} = E_{\text{sec}}/E_{\text{proj}}$. Only high-energy secondaries with $x_{\text{lab}} \ge x_{\text{lab}}^{\text{thr}}$ are enhanced, motivated by the requirement for an efficient mechanism that reduces the fraction of energy transferred into the electromagnetic component via $\pi^0$ production.

The enhancement factor $\zeta$ is defined as
\begin{align}
  \zeta(E_{\rm proj}, &x_{\rm lab}) =
  \Theta(E_{\rm proj}-E_{\rm start})\, \Theta(x_{\text{lab}} - x_{\text{lab}}^\text{thr}) \nonumber \\
  &\times \min\!\left[1,\,
  \dfrac{\kappa}{100}\,
  \log_{10}\!\left(\tfrac{E_{\rm proj}}{E_{\rm start}}\right)\,
  \right],
\label{eq:swap_basic}
\end{align}
where $\kappa$ specifies the growth rate of $\zeta$ per energy decade and $E_{\text{start}}$ is the energy at which enhancement sets in. The second step function $\Theta(x_{\text{lab}} - x_{\text{lab}}^\text{thr})$ restricts the effect to secondaries with $x_{\text{lab}}$ above the threshold $x_{\text{lab}}^{\text{thr}}$. Thus, Eq.~\ref{eq:swap_basic} fully defines the model in terms of three parameters: $\kappa$, $E_{\text{start}}$, and $x_{\text{lab}}^{\text{thr}}$.

The enhancement factor $\zeta$ is inferred from measurements of the $K/\pi$ production ratio in accelerator experiments. Since pions are produced much more abundantly than kaons, a small enhancement factor $\zeta$ can result in a relatively large change in the kaon yield. This amplification effect means that the precision achievable for $\zeta$ can be better than the precision of the $K/\pi$ production ratio measurements. 
For example, if the ratio is measured with a precision of 8.4\% at FASER, it corresponds to a 2\% precision in $\zeta$. Similarly, if the ratio is measured with a precision of 10.8\% at LHCb, it corresponds to a 2.5\% precision of $\zeta$.

\subsection{Calculation of the Number of Muons}
The number of muons $N_\mu$ after applying strangeness enhancement is obtained from the derivative distribution $\mathcal{D}$ introduced in Sec.~\ref{sec:calculation_for_contour}, using parameter sets $(\kappa, E_{\text{start}}, x_{\text{lab}}^{\text{thr}})$ that define the enhancement function $\zeta(E_{\text{proj}}, x_{\text{lab}})$. The modified muon number is then obtained as 
\begin{align}
    N_\mu^{\star}(E_{\rm prim}) &= N_\mu(E_{\rm prim}) \bigg[ 1 + {} \nonumber \\
    &\quad \iint\limits_R \mathrm{d}x_{\rm lab}\,\mathrm{d}E_{\rm proj}\, \zeta(E_{\rm proj},x_{\rm lab}) \nonumber \\
    &\quad \times \mathcal{D}(E_{\rm prim},E_{\rm proj},x_{\rm lab}) \bigg] \,,
    \label{eq:Nmu-mod}
\end{align}
where $N_\mu(E_{\rm prim})$ is the unmodified muon number, and the integration region is defined by $R:\; x_{\text{lab}} \leq 1$, $0.1$~GeV $< E_{\text{proj}} < E_{\text{prim}}$, and $0.1$~GeV $< x_{\text{lab}} E_{\text{proj}} < E_{\text{prim}}$. 

In practice, the integral is evaluated by computing $\zeta\times\mathcal{D}$ in each bin and summing over the entire phase space. For bins near the $x_{\text{lab}}^{\text{thr}}$ boundary, each bin is subdivided into 100 finer bins, and $\zeta\times\mathcal{D}$ is averaged over them to improve numerical accuracy.

For a primary nucleus of mass number $A$, we adopt the superposition model of nucleon interactions, which is appropriate for calculating means of air shower observables. The muon number is then given by distributing the primary energy among $A$ nucleons and summing their contributions:
\begin{align}
    N_\mu^{\star,A}(E_{\rm prim}) &= A\,N_\mu\!\bigl(\tfrac{E_{\rm prim}}{A}\bigr) \bigg[ 1 + {} \nonumber \\ 
    &\quad \iint\limits_{R'} \mathrm{d}x_{\rm lab}\,\mathrm{d}E_{\rm proj}\, \zeta(E_{\rm proj},x_{\rm lab}) \nonumber \\
    &\quad \times \mathcal{D}^A(E_{\rm prim},E_{\rm proj},x_{\rm lab}) \bigg] \,,
    \label{eq:Nmu_from_derivative_massA}
\end{align}
with 
\begin{equation}
    \mathcal{D}^{A}(E_{\text{prim}}, E_{\text{proj}}, x_{\text{lab}}) 
    = A\, \mathcal{D}^{\text{proton}}\!\Big(\tfrac{E_{\text{prim}}}{A},\, E_{\text{proj}},\, x_{\text{lab}}\Big)\,,
\end{equation}
where $\mathcal{D}^{\text{proton}}$ is the derivative matrix for a proton primary. The integration region $R'$ is defined as $x_{\text{lab}} \leq 1$, $0.1$~GeV $< E_{\text{proj}} < E_{\text{prim}}/A$, and $0.1$~GeV $< x_{\text{lab}} E_{\text{proj}} < E_{\text{prim}}/A$. 

To calculate the measured number of muons, the composition of the primary cosmic rays has to be considered. The PAO measurements~\cite{AugerComposition2014,AugerComposition2017} provide energy-dependent fractional abundances $F_j(E)$ assuming four composition groups, proton, helium, nitrogen, and iron. We use this four composition model to calculate the number of muons at the ground.
The total muon number for a mixed composition is given by
\begin{equation}
    N_\mu^*(E_{\text{prim}}) = \sum_j F_j\, N_{\mu,j}^*(E_{\text{prim}})\,,
    \label{eq:Nmu-mixed}
\end{equation}
where $j$ denotes the composition groups (p, He, N, Fe), and $F_j$ is the fractional abundance of each group. 
We adopt the measurements by the PAO~\cite{AugerComposition2014,AugerComposition2017,PierreAuger:2023xfc}, interpolate to the relevant energies, and normalize to unity. 
In this study, we use the composition fractions from Refs.~\cite{AugerComposition2014,AugerComposition2017}, which provide the energy-dependent $F_j(E)$ values above $10^{8.3}~\mathrm{GeV}$. 

For the calculation with $E_{\rm prim}=10^7, 10^8~\mathrm{GeV}$, the GSF 2025 composition~\cite{Dembinski:2025nmp} is used with the composition groups (p, He, O, Fe), where oxygen is used instead of nitrogen to represents a carbon-nitrogen-oxygen light nuclei group. 

The muon numbers obtained with this method are consistent with the recent study using the AIRES simulation package~\cite{Anchordoqui:2022fpn}. Implementing the toy model of Sec.~II in Ref.~\cite{Anchordoqui:2022fpn}, we find
\begin{align}
\langle R_\mu \rangle &= 1.86 \quad \text{for} \quad E_{\rm prim} = 10^{10}\,\mathrm{GeV},\; f_s = 0.4\;, \\
\langle R_\mu \rangle &= 2.07 \quad \text{for} \quad E_{\rm prim} = 10^{10}\,\mathrm{GeV},\; f_s = 0.6\;,
\end{align}
where $\langle R_\mu \rangle$ is the muon yield normalized to $1.455\times 10^7$. These values agree with Fig.~3 of Ref.~\cite{Anchordoqui:2022fpn} within 10\%.

\subsection{Likelihood Function for Parameter Constraints}
\label{sec:method_for_likelihood}
We define a likelihood function $\mathcal{L}$ to constrain the scenario parameters using both air-shower and accelerator data. The muon number measured by the PAO constrains $N_\mu^*$ as given in Eq.~\ref{eq:Nmu-mixed}, while accelerator experiments constrain the enhancement factor $\zeta$ through pion and kaon production measurements relative to the hadronic interaction model employed to obtain the $N_\mu^*$ expectation. The likelihood is given by
\begin{align}
    \mathcal{L} &= \frac{1}{\sqrt{2\pi}\,\sigma^{\rm PAO}} \nonumber \\
    &\quad \times \exp\!\Biggl[-\frac{1}{2}\Bigl(\frac{N_\mu(\kappa,E_{\rm start},x_{\rm lab}^{\rm thr})
      - N_\mu^{\rm PAO}}{\sigma^{\rm PAO}}\Bigr)^{\!2}\,\Biggr] \nonumber \\
    &\quad \times \prod_i \frac{1}{\sqrt{2\pi}\,\sigma_i^{\rm acc}} \nonumber \\
    &\quad \times \exp\!\Biggl[-\frac{1}{2}\Bigl(\frac{\zeta_i(\kappa,E_{\rm start},x_{\rm lab}^{\rm thr})
      - \zeta_i^{\rm acc}}{\sigma_i^{\rm acc}}\Bigr)^{\!2}\,\Biggr],
    \label{eq:likelihood}
\end{align}
where $N_\mu^{\text{PAO}}$ and $\sigma_{\text{PAO}}$ denote the central value and total uncertainty (including systematics) of the PAO muon measurement, and $\zeta_i^{\text{acc}}$ with $\sigma_i^{\text{acc}}$ are the enhancement values and uncertainties from accelerator data. The index $i$ runs over all relevant data points from LHC experiments.

The three free parameters $(\kappa, E_{\text{start}}, x_{\text{lab}}^{\text{thr}})$ are fitted by maximizing $\mathcal{L}$ using the \textsc{iminuit} package~\cite{iminuit}. To quantify parameter constraints, we use the likelihood-ratio test statistic
\begin{equation}
    \lambda = -2\Big[\ln \mathcal{L}(\kappa, E_{\text{start}}, x_{\text{lab}}^{\text{thr}}) - \ln \mathcal{L}_{\max}(\hat{\kappa}, \hat{E}_{\text{start}}, \hat{x}_{\text{lab}}^{\text{thr}})\Big]\,,
\end{equation}
where $\mathcal{L}_{\max}$ is the maximum likelihood, obtained at the best-fit parameters $\hat{\kappa}$, $\hat{E}_{\text{start}}$, and $\hat{x}_{\text{lab}}^{\text{thr}}$. 

\section{Experimental and Observational Constraints on Strangeness Enhancement}
\label{sec:results_constraints}

\subsection{A Strangeness Enhancement Scenario for Explaining the PAO Result}

The simulation framework developed here translates experimental observables into constraints on the strangeness enhancement scenario. This facilitates a quantitative assessment of the model's viability and determines the specific parameter configurations that can be constrained or excluded by upcoming data.

\begin{figure}
  \centering
  \includegraphics[width=0.5\textwidth]{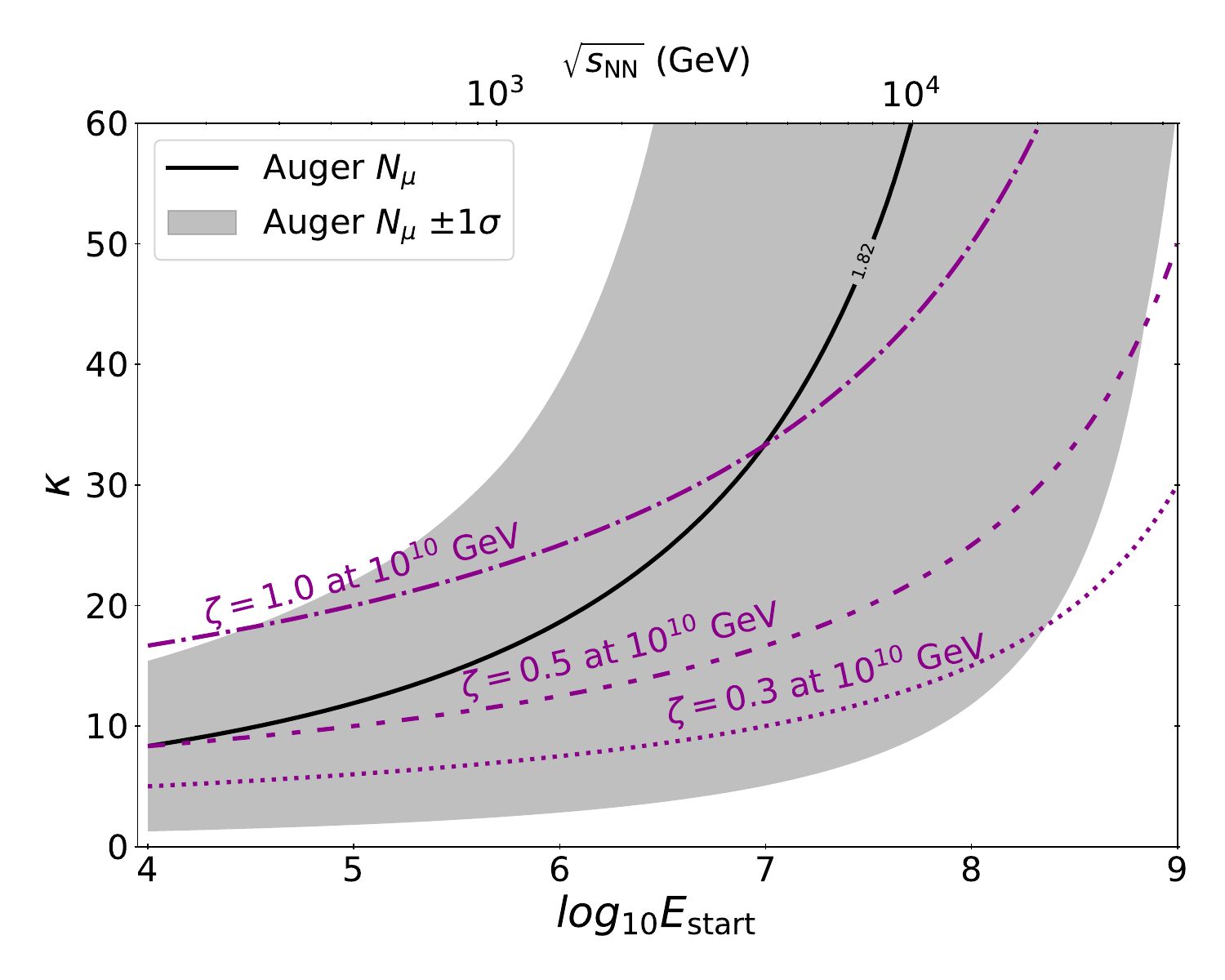}
  \caption{Allowed region in the $(\kappa, E_{\text{start}})$ plane, where the parameters $\kappa$ and $E_{\text{start}}$ are defined in Eq.~\ref{eq:swap_basic}, that reproduces the PAO muon count. The solid line corresponds to the PAO central value (for $x_{\text{lab}}^{\text{thr}} = 10^{-12}$, effectively no threshold). The grey-hatched band indicates the $\pm 1\sigma$ range. Dash-dotted, dash–double-dotted, and dotted lines mark where $\zeta$ reaches 1.0, 0.5, and 0.3, respectively, at $10^{10}$~GeV.}
  \label{fig:param_space}
\end{figure}

First, using calculated muon numbers $N_\mu^* (E_{\text{prim}})$, we scanned the parameter space to identify parameter values that reproduce the central value of the number of muons, as well as its $\pm 1\sigma$ range. The solid line in Fig.~\ref{fig:param_space} indicates the combinations of $(\kappa, E_{\text{start}})$ (for a very low threshold $x_{\text{lab}}^{\text{thr}} = 10^{-12}$, chosen to effectively eliminate the threshold cut) that yield the central value of the PAO muon count. The grey-hatched region corresponds to the $\pm 1\sigma$ band around this measurement. This band includes extreme cases where $\zeta$ reaches 1 (all pions swapped with kaons) at or below $10^{10}$~GeV. The dash-dotted, dash–double-dotted, and dotted lines in Fig.~\ref{fig:param_space} indicate where $\zeta$ reaches 1.0, 0.5, and 0.3 at $10^{10}$~GeV, respectively. Although a value of $\zeta > 0.5$ at $10^{10}$~GeV is needed to match the central PAO value, values below 0.3 are sufficient to remain within the $1\sigma$ range.

\begin{figure}
    \centering
    \includegraphics[width=0.5\textwidth]{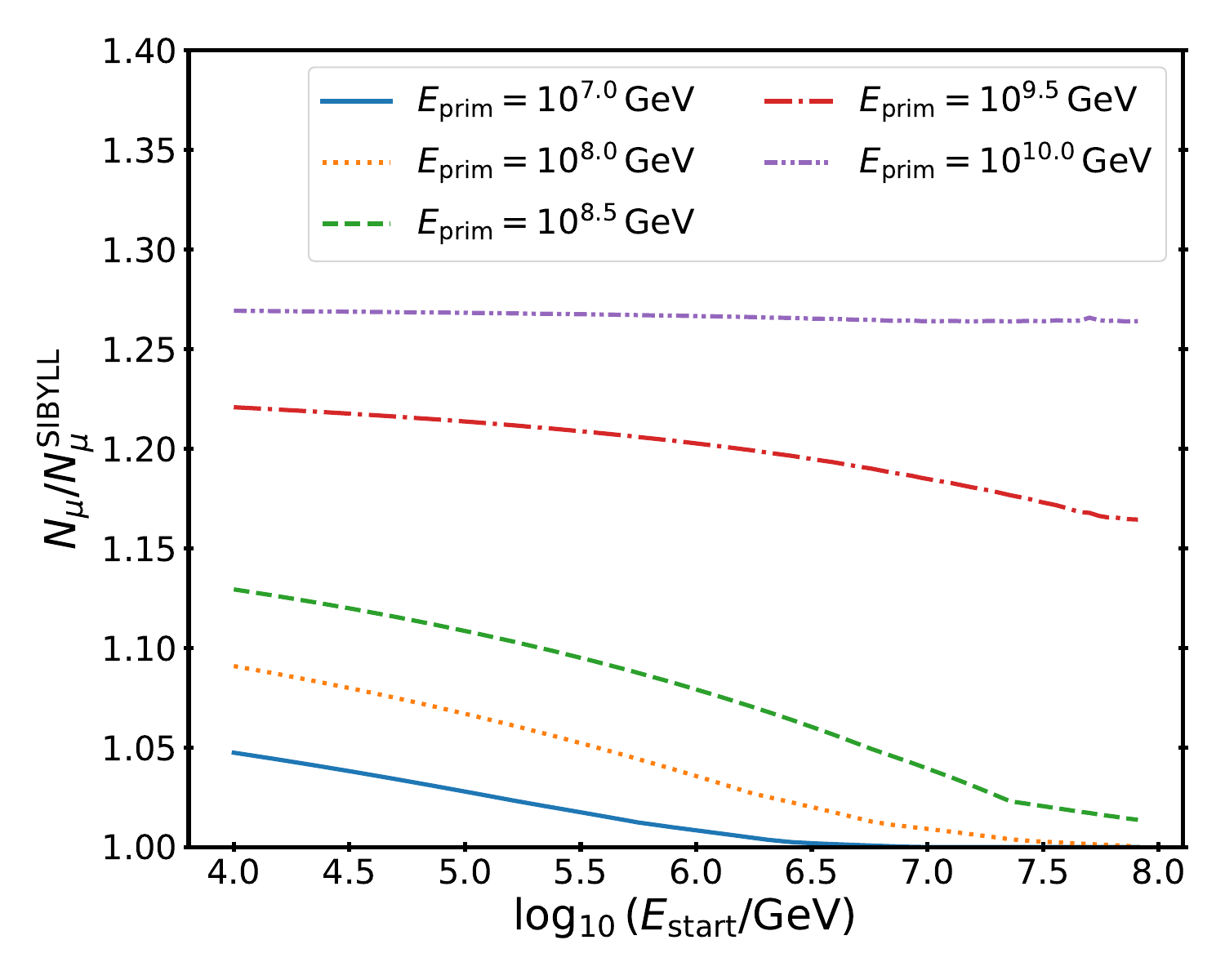}
    \caption{The ratio of the number of muons with respect to the original prediction by SIBYLL~2.3d. The parameter sets reproducing the central PAO value are applied to emulate the strangeness enhancement. 
    The horizontal axis shows $\log_{10}(E_{\rm start}/\mathrm{GeV})$ value from the parameter sets that reproduce the central value of the PAO measurement.
    The ratio is calculated for five different primary cosmic-ray energies, $10^{7}$ (solid line), $10^{8}$ (dotted line), $10^{8.5}$ (dashed line), $10^{9.5}$ (dash-dotted line) and $10^{10}~\mathrm{GeV}$ (dash-two-dotted line). The composition fractions from the Auger experiment~\cite{AugerComposition2017} are used for $10^{8.5}$, $10^{9.5}$, and $10^{10}~\mathrm{GeV}$, while the composition fractions by the GSF 2025~\cite{Dembinski:2025nmp} are used for $10^{7}$ and $10^{8}~\mathrm{GeV}$}
    \label{fig:ratio_Nmu_auger_center}
\end{figure}

Second, the viable parameter space must be consistent with the energy dependence of the observed muon excess. Using the parameter combinations that reproduce the PAO muon count at $10^{10}~\mathrm{GeV}$ (Fig.~\ref{fig:param_space}), Eq.~\ref{eq:Nmu-mixed} yields the expected number of muons at lower primary energies. Figure~\ref{fig:ratio_Nmu_auger_center} illustrates the ratio of this modified muon count to the SIBYLL~2.3d baseline prediction as a function of the onset energy $E_{\rm start}$. The curves follow the enhancement at primary energies of $10^{7}$, $10^{8}$, $10^{8.5}$, and $10^{9.5}~\mathrm{GeV}$, scaled to ensure a 27\% boost at the highest energy. Recent meta-analyses~\cite{WHISP2023} show the excess emerging around $10^{8}~\mathrm{GeV}$ and growing with primary energy, to explain the PAO measurements at $10^{8.5}$ and $10^{9.5}~\mathrm{GeV}$~\cite{PierreAuger:2020gxz}. Although a direct comparison is limited because \textsc{MCEq} simulates longitudinal developments rather than lateral ground distributions, capturing the general trend requires that the modified-to-baseline ratio remains near 1.0 at $10^{7}~\mathrm{GeV}$, rises slightly above unity by $10^{8}~\mathrm{GeV}$, and increases systematically at higher energies. These physical conditions tightly constrain the onset energy $E_{\rm start}$ to the range of $10^{6}-10^{7}~\mathrm{GeV}$.

For this favored configuration, the required magnitude of strangeness enhancement at typical LHC energies is substantial. Assuming an onset of $E_{\rm start}=10^6~\mathrm{GeV}$, the enhancement parameter $\zeta$ evaluates to 0.375. This modification effectively replaces over one-third of the relevant pion yield with kaons, nearly doubling the overall inclusive kaon production compared to the SIBYLL~2.3d baseline. Such a pronounced shift offers a distinct phenomenological signature, which upcoming measurements of the inclusive pion-to-kaon ratio at forward-physics experiments like LHCb, FASER, and SND@LHC can directly test.

\subsection{Adding Constraints from Collider Experiments}

While reproducing the cosmic-ray measurements requires relatively large enhancement factors, there remains a substantial possibility that LHC experiments will observe no such effect. To evaluate the requisite sensitivity for collider experiments to constrain the model, we calculate the parameter space that would be rejected under the null hypothesis of zero observed enhancement at a given measurement precision. Measurements in the forward region ($x_{\rm lab} \sim 0.1$--$0.2$) are particularly effective, as they probe the phase space where the predicted enhancement effects are most pronounced.

\begin{figure}
    \centering
  \includegraphics[width=0.98\linewidth]{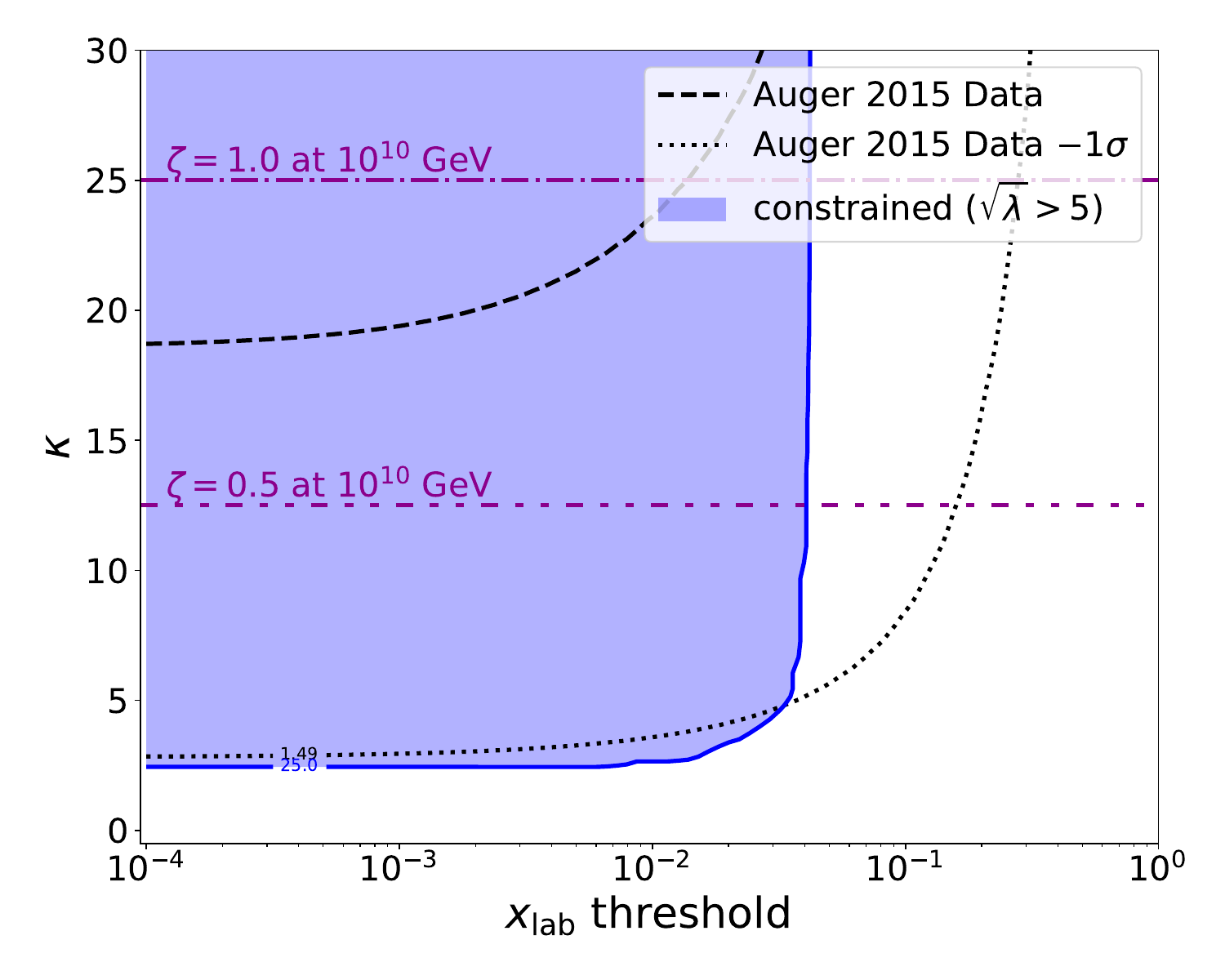}  
    \caption{Expected constraint from a projected LHCb Run~3 measurement (hatched region) assuming no observed enhancement at 2.5\% precision ($\zeta_{\text{LHCb}} = 0.00 \pm 0.025$), calculated for $E_{\text{start}} = 10^{6}$~GeV. The black dashed and dotted lines are the PAO central value (dashed) and $\pm 1\sigma$ range (dotted) of the Auger muon measurement (same as in Fig.~\ref{fig:param_space}). Magenta dash-dotted and dash–double-dotted lines indicate $\zeta = 1.0$ and $0.5$ at $10^{10}$~GeV, respectively.}
    \label{fig:LHCb_constraint}
\end{figure}

\begin{figure}
    \centering
  \includegraphics[width=0.98\linewidth]{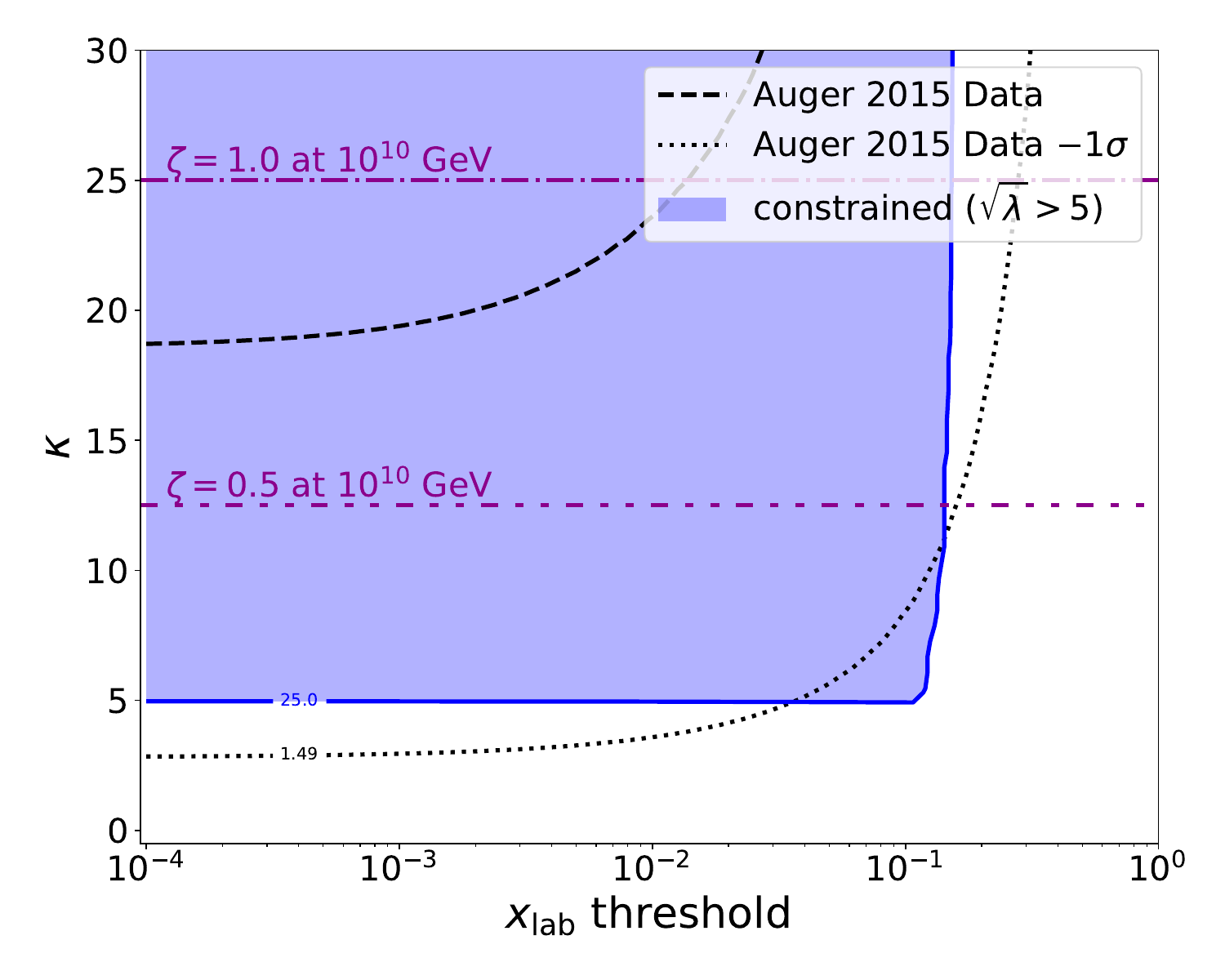} 
    \caption{Expected constraint from a projected FASER Run~3 measurement (hatched region) assuming no observed enhancement at 2\% precision ($\zeta_{\text{FASER}} = 0.00 \pm 0.02$), calculated for $E_{\text{start}} = 10^{6}$~GeV. The black dashed and dotted lines are the same PAO reference curves as in Fig.~\ref{fig:param_space}.}
    \label{fig:FASER_constraint}
\end{figure}

Our projections utilize the reported PAO muon measurement alongside anticipated Run~3 data from LHCb and FASER. While SND@LHC~\cite{SNDLHC:2022ihg} also provides neutrino-based measurements with flavor separation, FASER is positioned directly along the collision axis. This location captures a highly collimated neutrino flux from forward $\pi^\pm$ and $K$ decays~\cite{FASERneutrinoFlux2024,Abbaneo:2926288}, yielding higher expected statistics for tighter constraints. Consequently, SND@LHC is considered a complementary cross-check for future analyses. For the LHCb projection, five pseudorapidity bins from Run~2 charged-particle measurements~\cite{LHCbRun2} were converted to $x_{\text{lab}}$. For FASER, a single $x_{\text{lab}}$ bin was defined based on detector coverage and expected Run~3 neutrino yields (see Appendix~A). Consistent with the favored energy dependence derived earlier, we assume $E_{\rm start} = 10^{6}~\mathrm{GeV}$ and define the constraint boundaries at a contour level of $\sqrt{\lambda} = 5$.


Assuming LHCb finds no strangeness enhancement, with a precision of about 2.5\% (i.e. $\zeta_{\text{LHCb}} = 0.00 \pm 0.025$), we computed the excluded regions of parameter space using the likelihood method of Sec.~\ref{sec:method_for_likelihood}. Figure~\ref{fig:LHCb_constraint} shows the resulting $5\sigma$ exclusion region as the blue-hatched area, calculated for $E_{\text{start}} = 10^{6}$~GeV. The black dashed and dotted lines denote the PAO central value and $\pm 1\sigma$ range for $N_\mu$. We find that for $x_{\text{lab}}^{\text{thr}} < 3\times 10^{-2}$, most of the parameter space compatible with the PAO measurement is excluded by the hypothetical LHCb result. Higher values of $x_{\text{lab}}^{\text{thr}}$ remain allowed due to the limited forward acceptance of the LHCb detector.

\begin{figure}
    \centering
  \includegraphics[width=0.98\linewidth]{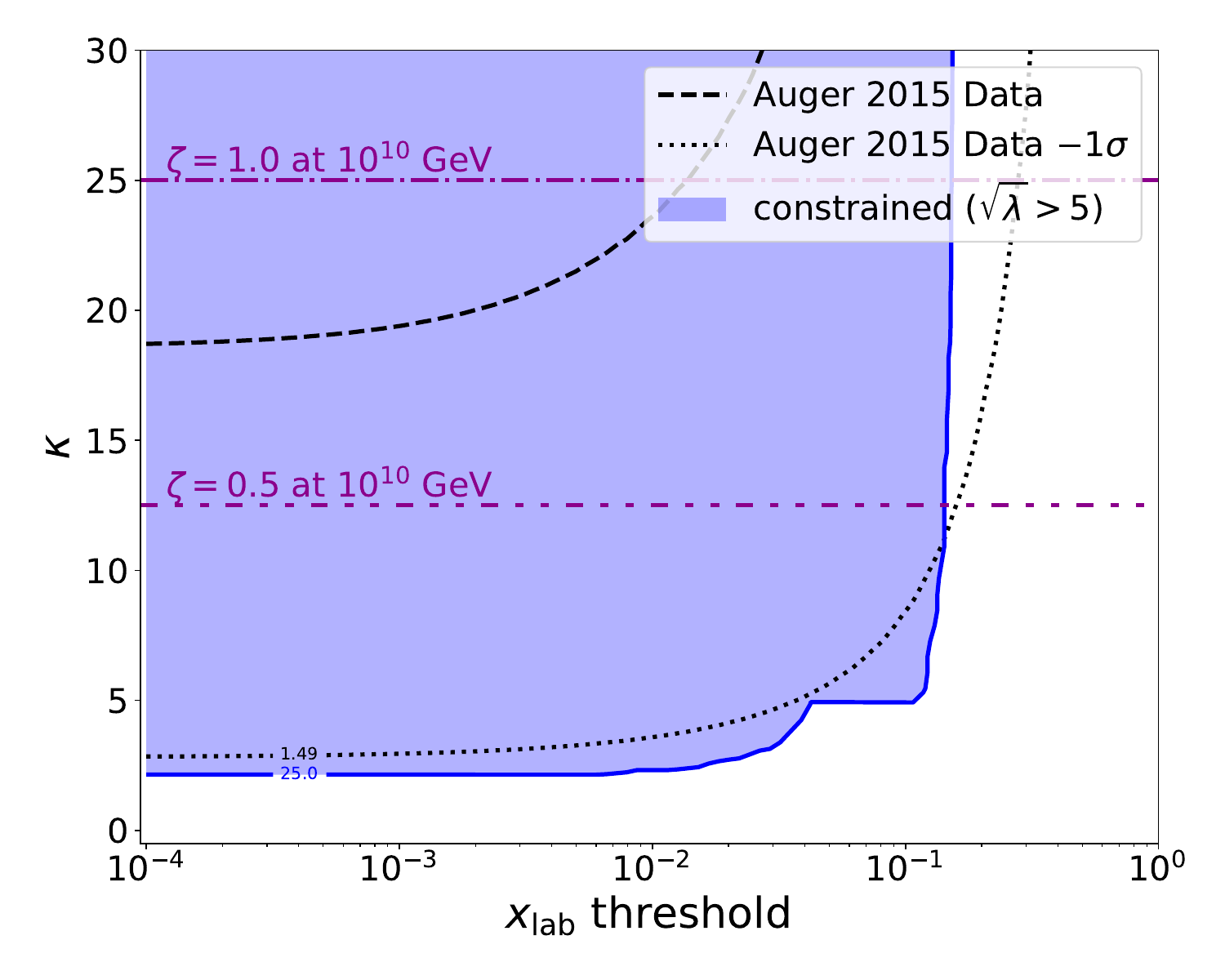}
    \caption{Combined constraints from projected LHCb (2.5\% precision) and FASER (2\% precision), assuming neither experiment observes an enhancement. Nearly the entire parameter space consistent with the Auger muon excess (grey band) is excluded (blue hatching), except for scenarios with very high $x_{\text{lab}}^{\text{thr}}$, calculated for $E_{\text{start}} = 10^{6}$~GeV.}
    \label{fig:combined_constraint}
\end{figure}
Assuming no observed enhancement at FASER, with about 2\% precision ($\zeta_{\text{FASER}} = 0.00 \pm 0.02$), Fig.~\ref{fig:FASER_constraint} shows the analogous exclusions from FASER data alone. The LHCb-based exclusion (from Fig.~\ref{fig:LHCb_constraint}) is less restrictive in the high-$x_{\text{lab}}^{\text{thr}}$ region because LHCb has no coverage above $x_{\text{lab}}\sim 0.05$. By contrast, FASER’s neutrino-based measurement covers a broader range in $x_{\text{lab}}$, excluding all parameter space with $x_{\text{lab}}^{\text{thr}} \lesssim 0.1$. Nevertheless, some low-$\zeta$ regions remain allowed: regions with $\kappa \lesssim 5.0$ (max $\zeta \approx 0.2$) by FASER alone.

Finally, assuming that neither LHCb nor FASER observes any strangeness enhancement (at 2.5\% and 2\% precision, respectively), Fig.~\ref{fig:combined_constraint} shows the combined excluded region (blue hatching) at $5\sigma$. In this combined scenario, nearly all parameter space within the $\pm 1\sigma$ range of the PAO muon result can be ruled out. Only scenarios with very high $x_{\text{lab}}^{\text{thr}}$ ($> 2\times 10^{-1}$) remain allowed. These correspond to cases where the enhancement manifests only for extremely high-$x_{\text{lab}}$ secondaries, requiring a larger $\zeta$ value at the highest energies to still produce the Auger-observed muon excess. For instance, the remaining region corresponds to $\zeta > 0.4$ at $10^{10}$~GeV. We also find that achieving a $\zeta$ precision of 2.5\% at LHCb would require measuring the $K/\pi$ ratio to better than $10.8\%$, while achieving a $\zeta$ precision below 2\% at FASER would require better than $8.4\%$ precision.

\section{Conclusions\label{sec:conclusion}}
Ultra-high-energy cosmic-ray air showers produce more muons than predicted by current simulations, highlighting the long-standing muon puzzle. One of the proposed explanations is a scenario in which muon production is enhanced through an increased yield of strange particles at high projectile energies and large energy fractions. However, because accelerator experiments cover only a limited region of the relevant phase space, such a scenario cannot be tested using accelerator data alone. In this work, we have shown that combining accelerator and air-shower measurements within a model-based framework provides a powerful strategy for testing this hypothesis.

We mapped the interaction phase space relevant to muon production and quantified the sensitivity of the muon yield to strangeness enhancement by modifying particle production in \textsc{MCEq} simulations. Our analysis identified hadronic interactions with projectile energies above $10^4$~GeV and secondary particles with $x_{\text{lab}} > 10^{-3}$ as the most critical for altering the muon content of air showers.

Our calculation framework allowed us to connect the model expectations and measurements from both cosmic-ray and accelerator experiments. Using parameter sets that reproduce the PAO measurement, the expected number of muons at $10^{7}$, $10^{8}$, $10^{8.5}$, and $10^{9.5}~\mathrm{GeV}$ was evaluated. Observations from cosmic-ray experiments~\cite{WHISP2023,PierreAuger:2020gxz,Albrecht:2021cxw} show that the muon excess emerges around $10^{8}~\mathrm{GeV}$ and is clearly present at $10^{8.5}$ and $10^{9.5}~\mathrm{GeV}$. To match this energy dependence, parameter sets with an onset energy $E_{\rm start}$ of $10^{6}-10^{7}~\mathrm{GeV}$ are favored. For $E_{\rm start}=10^{6}~\mathrm{GeV}$, the required strangeness enhancement is $\zeta=0.375$, which can be directly tested by measurements at the LHC.

Conversely, it is entirely possible that no such enhancement will be observed at the LHC. Therefore, the precision required for LHC experiments to successfully reject the parameters reproducing the PAO measurements was also evaluated.
Assuming no observed enhancement, the projected exclusion regions show that reproducing the PAO muon excess within its $\pm1\sigma$ range would be incompatible with the combined accelerator constraints if $\zeta$ can be measured with about 2.5\% precision at LHCb and better than 2\% at FASER. These requirements translate into approximately 10.8\% and 8.4\% precision, respectively, on the $K/\pi$ production ratio.

Our results demonstrate that meaningful experimental tests of the strangeness enhancement scenario are achievable despite current limitations in phase space coverage. The forthcoming LHC Run~3 data will provide the first direct constraints on this hypothesis and represent an important step toward resolving the muon puzzle. The calculation framework developed here can be extended to other scenarios for the muon puzzle. 

\begin{acknowledgements}
AF acknowledges support from Academia Sinica (Grant No.~AS-GCS-113-M04) and the National Science and Technology Council (Grant No.~113-2112-M-001-060-MY3).
\end{acknowledgements}

\appendix

\section{Accelerator experiments for constraining the strangeness enhancement\label{sec:app:experiment_summary}}

LHC experiments probe the phase space relevant to the muon puzzle to $\sqrt{s} \leq 1.4\times10^4~\mathrm{GeV}$. In Sec.~\ref{sec:results_contour}, we identified interactions with $E_{\rm proj}>10^4$~GeV and $x_{\rm lab}>10^{-3}$ as critical. Measurements of pion and kaon production here provide essential input. ATLAS, CMS, and ALICE cover only central rapidities, while LHCb extends to $2<\eta<5$ and can identify charged pions and kaons up to $x_{\rm lab}<0.05$~\cite{Albrecht:2021cxw,LHCbRun2}. Accessing $x_{\rm lab}>0.05$ requires far-forward detectors, which generally lack dedicated particle identification capabilities.  

Forward experiments constrain pion and kaon production indirectly through decay products. LHCf measures photons from $\pi^0$ decays and reconstructs $\pi^0\to\gamma\gamma$ for neutral pions above 20\% of the beam energy~\cite{LHCfpi0_ICRC2021,LHCfpi0_2015}. FASER~\cite{FASER_detector} and SND@LHC~\cite{SNDLHC:2022ihg} detect neutrinos from pion and kaon decays, with flavor sensitivity providing separation of pion and kaon contributions.  
The method relies on comparing the yields of electron neutrinos ($\nu_e + \bar{\nu}_e$) and muon neutrinos ($\nu_\mu + \bar{\nu}_\mu$), which have different origins: muon neutrinos arise from pion, kaon, and charm hadron decays, while electron neutrinos originate predominantly from kaon and charm hadron decays. To isolate the pion and kaon contributions, we restrict the analysis to low-energy neutrinos where charm hadron decay contributions are negligible. Specifically, for FASER, we consider neutrinos below 1~TeV for $\nu_\mu$ and below 700~GeV for $\nu_e$, where charm meson decays do not dominate~\cite{FASERneutrinoFlux2024}. In this energy range, when charm hadron contributions are negligible, the ratio $\nu_e/\nu_\mu$ approximately reflects the ratio of kaon production to the sum of pion and kaon production, $R_{K/(K+\pi)} \approx N_{\nu_e}/N_{\nu_\mu}$.

Based on the expected Run~3 statistics from EPOS-LHC predictions~\cite{FASERneutrinoFlux2024}, we anticipate $\mathcal{O}(300)$ detected $\nu_e$ events and $\mathcal{O}(2400)$ detected $\nu_\mu$ events (assuming a 30\% detection efficiency), corresponding to statistical uncertainties of approximately 5--7\%. This precision is sufficient to constrain the enhancement factor $\zeta$ and test the strangeness enhancement scenario. SND@LHC~\cite{SNDLHC:2022ihg} also provides complementary measurements with neutrino flavor separation, and its expected Run~3 statistics are around 2000. We use FASER as the baseline case because of its projected statistics in Run 3; SND@LHC provides an essential, independent cross-check with different acceptance and systematics.

Detector acceptance for pions and kaons in FASER and SND@LHC is estimated using Rivet~\cite{Kling:2021gos} at $\sqrt{s}=13$~TeV. 
FASER probes kaons in $0.06<x_{\rm lab}<0.20$ and pions in $0.04<x_{\rm lab}<0.20$, using $\nu_e$ and $\nu_\mu$. SND@LHC covers narrower ranges ($x_{\rm lab}\sim0.02\mbox{--}0.14$). 
In Sec.~\ref{sec:results_constraints}, we used projected LHC Run~3 data: five pseudorapidity bins from LHCb (Table~\ref{tab:LHCb_bin}) and a single bin $0.06<x_{\rm lab}<0.20$ for FASER.  

\begin{table}[ht]
  \centering
  \begin{tabular}{c|cc}
    \hline
    Bin & Pseudorapidity range & Maximum $p_T$ \\
    \hline\hline
    1 & 4.5--4.9 & 5~GeV \\
    2 & 4.0--4.5 & 7~GeV \\
    3 & 3.5--4.0 & 8~GeV \\
    4 & 3.0--3.5 & 10~GeV \\
    5 & 2.5--3.0 & 10~GeV \\
    \hline
  \end{tabular}
  \caption{Projected LHCb Run~3 bins for charged-particle multiplicity.}
  \label{tab:LHCb_bin}
\end{table}

\section{Measurements of the number of muons}

PAO reported an excess in the number of muons relative to simulation predictions~\cite{MuonAuger2015}. They measured the mean logarithm of the muon content at $10^{19}~\mathrm{eV}$,
\begin{equation}
  \langle\ln R_\mu\rangle(10^{19}~\mathrm{eV}) = 0.601 \pm 0.016\ ^{+0.167}_{-0.201}\ (\mathrm{sys}),
\end{equation}
where $R_\mu$ denotes the number of muons normalized to $N_\mu=1.455\times10^7$. We adopt this value for comparison in our analysis.

\bibliographystyle{spphys}       
\bibliography{references}   

\end{document}